\documentclass[%
 manuscript=article,
 reprint,
 amsmath,amssymb,
 aps,
 prb,
]{revtex4-1}

\usepackage{chemformula} 
\usepackage[T1]{fontenc} 
\usepackage{graphics}
\usepackage{makecell}
\usepackage{xcolor}
\usepackage{amsmath}
\usepackage{enumerate}
\usepackage[normalem]{ulem}
\usepackage{multirow}
\usepackage[bottom]{footmisc}
\usepackage{bm}
\usepackage{siunitx}
\usepackage{multirow}
\begin{document}
\newcommand*\mycommand[1]{\texttt{\emph{#1}}}
\newcommand*\rd[1]{{\color{red} #1}}
\newcommand*\bl[1]{{\color{blue} #1}}
\newcommand\mathplus{+}
\newcommand{\snrxn}[0]{S$_\mathrm{N}$2}
\newcommand{\erxn}[0]{E2}
\newcommand{\Ea}[0]{$E_\mathrm{a}$}
\newcommand{\dEa}[0]{$\Delta E_\mathrm{a}$}
\newcommand{\Eas}[0]{$E_\mathrm{a}^{\mathrm{S}}$}
\newcommand{\Eae}[0]{$E_\mathrm{a}^{\mathrm{E}}$}
\newcommand{\deltaml}[0]{$\Delta$\nobreakdash-ML}

\title{Transition State Search and Geometry Relaxation throughout Chemical Compound Space with Quantum Machine Learning} 

\author{Stefan Heinen}
\affiliation{University of Vienna, Faculty of Physics, Kolingasse 14-16, AT-1090 Wien, Austria}
\author{Guido Falk von Rudorff}
\affiliation{University of Vienna, Faculty of Physics, Kolingasse 14-16, AT-1090 Wien, Austria}
\author{O. Anatole von Lilienfeld}
\email{anatole.vonlilienfeld@utoronto.ca}
\affiliation{Vector Institute for Artificial Intelligence, Toronto, ON, M5S 1M1, Canada}
\affiliation{Departments of Chemistry, Materials Science and Engineering, and Physics, University of Toronto, St. George Campus, Toronto, ON, Canada}
\affiliation{Machine Learning Group, Technische Universit\"at Berlin and Institute for the Foundations of Learning and Data, 10587 Berlin, Germany}
\begin{abstract}
We use energies and forces predicted within response operator based quantum machine learning (OQML) to perform  geometry optimization and transition state search calculations with legacy optimizers.
For randomly sampled initial coordinates of small organic query molecules we report systematic improvement of equilibrium and transition state geometry output as training set sizes increase.
Out-of-sample S$_\mathrm{N}$2 reactant complexes and transition state geometries have been predicted using the LBFGS and the  QST2 algorithm with an RMSD of 0.16 and 0.4 \r{A} --- after training on up to 200 reactant complexes relaxations and transition state search trajectories from the QMrxn20 data-set, respectively.
 For geometry optimizations, we have also considered relaxation paths up to 5'500 constitutional isomers with sum formula C$_7$H$_{10}$O$_2$ from the QM9-database. 
Using the resulting OQML models with an LBFGS optimizer reproduces the minimum geometry with an RMSD of 0.14~\r{A}.
For converged equilibrium and transition state geometries subsequent vibrational normal mode frequency analysis indicates deviation from MP2 reference results by on average 14 and 26\,cm$^{-1}$, respectively.
While the numerical cost for OQML predictions is negligible in comparison to DFT or MP2, the number of steps until convergence is typically larger in either case. 
The success rate for reaching convergence, however, improves systematically with training set size, underscoring OQML's potential for universal applicability.
\end{abstract}
\maketitle
%
%
\section{Introduction}
%
One of the fundamental challenges in quantum chemistry is the understanding of reaction mechanisms in order to predict chemical processes. 
To this end, numerous neural networks (reaction predictors) have been introduced, proposing the most likely reaction path way\cite{MLreact_rev_2020,Kayala2011, Wei2016, NIPS2017_6854, Fooshee2018, Segler2018, Schwaller2018} for a given product.
These models were trained on data obtained from experimental studies\cite{rxnfromlit} only containing the molecular graph (as SMILES strings\cite{smiles1, smiles2}) and their corresponding yields.
However, a crucial property of a chemical reaction is the activation energy (i.e. the difference between reactant and transition state energy), linked to the kinetics of the reaction.
To predict activation energies with conventional electronic structure methods, both the reactant complex geometry and the transition state geometry need to be obtained. This is commonly done by iteratively following gradients of the potential energy surface (PES) towards the minimum or the saddle point, respectively. 
Due to the iterative nature of these schemes, imposing the repeated need to perform self-consistent field calculations to obtain updated forces, the computational burden is as large as it is predictable\cite{Heinen_2020_cost_of_qm}. 
Furthermore, finding saddle points remain an additional challenge because often enough considerable manual work is required beforehand in order to generate reasonable initial structure guesses. 
Consequently, it is not surprising that so far only  few reaction data-sets  which contain transition state geometries as well as corresponding energies have been published in 2020~\cite{QMrxn20, grambow2020}, and 2021~\cite{Jackson_2021}.

Only very recently, attempts have been made to use machine learning models  to speed up transition state predictions.
In 2019, Bligaard and co-workers used the nudged elastic band (NEB)\cite{Henkelman_2000, henkelman2000improved} method  to find transition states relying on neural network based $\Delta$-ML model~\cite{deltaML2015} 
together with a low level of theory as baseline\cite{NEB_ml}.
More recently, Mortensen et al.~contributed the `atomistic structure learning algorithm' (ASLA)\cite{asla}, enabling autonomous structure determination with much reduced need for costly first-principles total energy calculations. 
Lemm \emph{et. al.}\cite{Lemm_2021} introduced the graph to structure (G2S) machine learning model,  predicting reactant complexes and transition state geometries for the QMrxn20\cite{QMrxn20} data-set without any account for energy considerations, solely using molecular graphs as input.
Also, for 30 small organic molecules neural networks predicting energies and forces to accelerate the geometry optimization in between \emph{ab initio} iterations was introduced by Meyer and Hauser~\cite{GPR_opt1}, and by Born and K\"astner~\cite{GPR_opt2}.
Similar to G2S, Mak\'os et al.~\cite{Mako__2021} propose a `transition state generative adversarial neural network' (TS-GAN) which estimates transition state geometries using information from reactants and products only. This procedure allows for better initial geometries for a transition state search reducing the number of steps towards a saddle point.
Jackson \emph{et. al.}\cite{Jackson_2021} developed a neural network (TSNet) predicting transition states for a small ($\sim$~50) S$_\mathrm{N}$2 reaction dateset, as well as geometries of the QM9\cite{qm9} data-set.
\begin{figure}[!tbp]
      \includegraphics[width=0.49\textwidth]{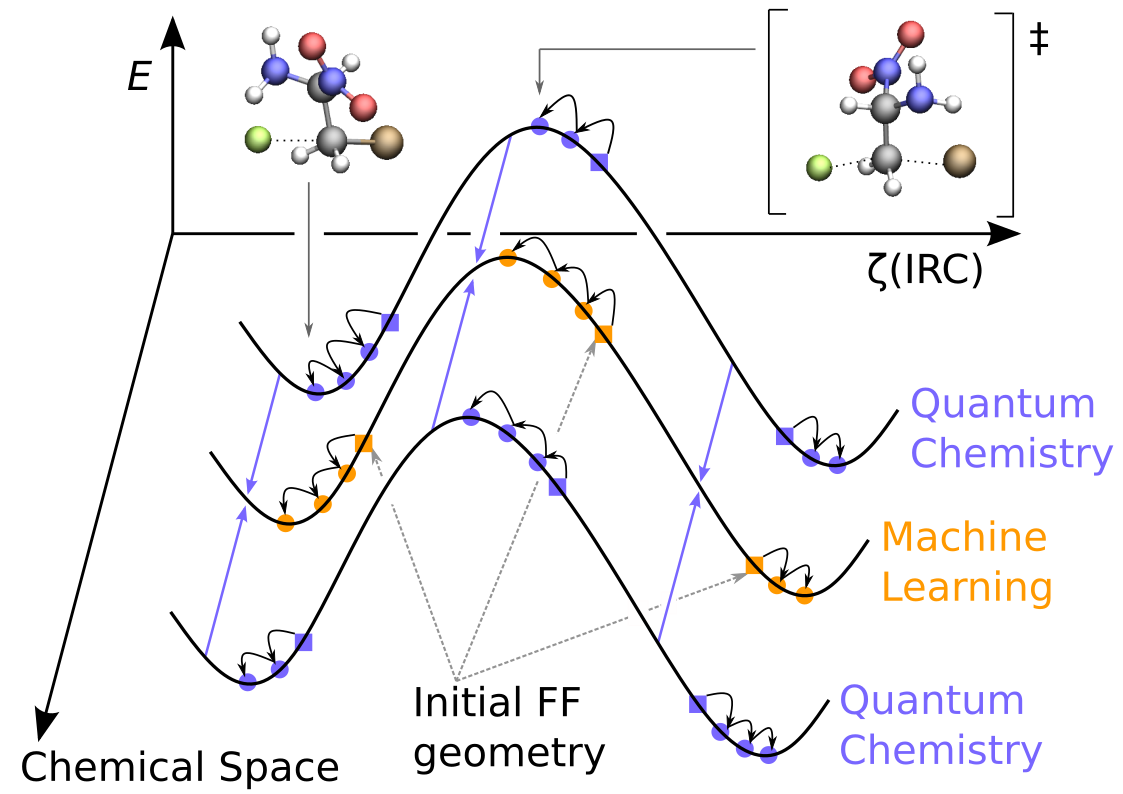}
  \caption{Schematic potential energy surfaces in chemical compound space. 
  Arrows show the working principle of OQML based iterative structural optimization:
  Training on reaction profiles of different chemical systems (purple), the OQML model is able to interpolate forces and energies throughout chemical compound space enabling the relaxation of the reactant and the search of the transition state (orange).
  Input geometries (squares) are easily obtained, e.g.~from universal force field predictions.
  }
  \label{fig:intro}
\end{figure}
%

However, to the best of our knowledge there is no machine learning model yet using, in strict analogy to the conventional quantum chemistry based protocol, predicted energies and forces only within the conventional optimization algorithms in order to relax geometries or find transition states.
To tackle this challenge, we have used for this paper the response operator based quantum machine learning (OQML)\cite{Christensen2019, Christensen2020} model with the FCHL representation~\cite{FCHL,FCHL19}
and trained on energies and forces across chemical compound space in order to speed up geometry relaxations as well as transition state searches
for new, out-of-sample compounds (see Fig.~1).  As for any properly trained QML model, prediction errors decay systematically with training set size, and we demonstrate for the chemistries presented that encouraging levels of accuracy can be reached. 

First, we have investigated geometry optimizations for all constitutional isomers with C$_7$H$_{10}$O$_2$ sum formula drawn from QM9\cite{qm9}.
After training the OQML model on random geometries along the optimization path of of 5500 calculations going from a UFF minimum energy geometry to the B3LYP/6-31G(2df) minimum geometry
we optimized the remaining 500 constitutional isomers resulting in a total RMSD of only 0.14 {\AA}.
To probe transition states, we have trained OQML models on the QMrxn20 data-set\cite{QMrxn20} with thousands of examples for the \snrxn\ text book reaction at MP2/6-311G(d) level of theory, enabling the relaxation of reactant complexes and the search of transition states which both compare well to common density functional theory (DFT) results. 
As shown in Figure \ref{fig:intro} this means training the OQML model on quantum chemistry reference energies and forces along the optimization trajectory obtained for relaxation and transition state search runs of training systems. Starting with universal force field (UFF)~\cite{Rappe1992} geometries, 
OQML subsequently predicts energies and forces for 200 out-of-sample query systems, thereby enabling the application of legacy relaxation and transition state search algorithms throughout chemical compound space. 

\section{Methods}
We have relied on operator quantum machine learning (OQML) approach as introduced by Christensen \emph{et. al.}\cite{Christensen2019, Christensen2020} which is a kernel ridge regression (KRR) model which explicitly encodes target functions and their derivatives.
A detailed derivation can be found in Christensen \emph{et. al.}\cite{FCHL19} section 2 (Operator quantum machine learning).
To train a model the regression coefficients $\bm{\alpha}$ the following cost function is minimized:
\begin{equation}
    J(\bm{\alpha}) = \begin{Vmatrix} \begin{bmatrix} \mathbf{y} \\ \mathbf{f} \end{bmatrix} - \begin{bmatrix} \mathbf{K} \\ -\frac{\partial}{\partial\mathbf{r}}\mathbf{K} \end{bmatrix} \bm{\alpha}  \end{Vmatrix}^2_2
    \label{eq:training}
\end{equation}
with $\mathbf{K}$ being the training kernel, $\mathbf{y}$ the energies, and $\mathbf{f}$ the forces.
To predict the energies following matrix equation can be used:
\begin{equation}
    \mathbf{y}^{\mathrm{est}} = \mathbf{K}_s \bm{\alpha}
\end{equation}
and similarly for the forces:
\begin{equation}
    \mathbf{f}^{\mathrm{est}} =  -\frac{\partial}{\partial\mathbf{r}}\mathbf{K}_s \bm{\alpha}
\end{equation}
where $\mathbf{K}_s$ being the test Kernel containing training and test instances.

The representation used throughout this work is the FCHL19 representation\cite{FCHL19}.
FCHL19 makes use of interatomic distances in its two body terms and includes interatomic angles in the three body term. 
FCHL19 was selected because of its remarkable performance for QM9 related data-sets\cite{FCHL19}, 
and due to being the best structure based representation in direct learning of activation energies in QMrxn20    ~\cite{Heinen2021}. 


To find transition states, the Gaussian09 QST2\cite{qst2} algorithm with loose convergence criteria was used, which allows for external energies and forces, in this case from OQML. Note that no explicit Hessian is required by this method nor is one available from our model.
For both, reactant complexes and transition states, 300 out of sample reactions were chosen.
For these 300 reactions also DFT geometry optimizations as well as transition state searches were performed.
The three functionals used were: B3LYP\cite{B3LYP_functional, B3LYP}, PBE0\cite{Adamo_1999}, and $\omega$B97X\cite{Mardirossian2014} with the 6-311G(d)\cite{binkley1980self,Petersson1988,Petersson1991} basis set (same as for the reference method in \cite{QMrxn20}).
More details of the training of the models can be found in the SI.

The python package rmsd\cite{rmsd_py} with the Kabsch algorithm\cite{Kabsch_1976} was used to obtain the RMSD's including hydrogens.
For every training set size, the success rate of the geometry optimization (truncated after 50 iterations) and the transition state search (truncated after 100 iterations which is the gaussian default) was reported.
Scripts and data can be found in the SI\cite{SI}.
\section{Results and Discussion}
In the context of statistical learning theory, cross-validated learning curves amount to numerical proof of the robustness and applicability of a machine learning model, and they provide quantitative measures of the data-efficiency obtained. 
For the three OQML models studied here-within (geometries of constitutional isomers, of reactant complexes, and of transition states), Fig.~\ref{fig:lc}(left) displays the OQML based learning curves for energies (top) and atomic forces (bottom) which indicate the systematic improvement of energy and force predictions as training set size increases.

\subsection{Geometry optimization}
The learning curve for the constitutional isomers are in line with the results by Christensen \emph{et. al.}\cite{Christensen2020}.
Surprisingly, although FCHL19 was optimized for small organic closed shell molecules, the learning curves for the reactant complexes and the transition states have a faster learning rate.
A possible reason for this trend could be that the reactions in the QMrxn20 data-set share a common scaffold with only the substituents changing which represents a lower effective dimensionality of the problem which typically leads to faster learning.
Also, relaxations for only 200 reactions were considered in the training set which implies an overall smaller subset of the chemical universe. 
By contrast, for the constitutional isomers, geometries from 5500 different compounds were chosen, covering a much broader chemical space.


While accurate OQML based estimates of forces and energies are necessary for subsequent relaxation and transition state search, the eventual key figure of merit, the RMSD with respect to query reference coordinates for increasing training set size, amounts to a performance curve as 
shown in the mid panel of Figure \ref{fig:lc}. 
We observe strong systematic improvements with increasing training set size of the RMSD for the constitutional isomers and reactant complexes. 
By contrast, RMSD performance curves for transition states, while also monotonically increasing with training set size, exhibit substantially smaller learning rates. 
Differences in learning for different data-sets while using the same representations and model architectures implies that  the target function is more complicated. 
One can argue that the constitutional isomers are less pathological since they consist of small organic and closed shell molecules, whereas the transition states include charged compounds and non-covalent binding to leaving and attacking groups.
The relatively flat progress made for the transition states might also simply due to the fact of the more complex optimization problems towards a saddle point compared to the simple downhill search of a geometry optimization. 
More specifically, due to the underpinning high dimensionality, the training set grows much more rapidly when adding a new reactive system including optimization steps along the way to the saddle point. This implies that the training will be less efficient. 
Possible ways to mitigate such a bottleneck could include the use of the Amons approach\cite{Amons} which decomposes molecules in sub-structures, drastically reducing the effective dimensionality of the problem.
Also $\Delta$-ML~\cite{deltaML2015}, multi-level grid combination techniques\cite{multilevel}, 
or transfer learning~\cite{transferL_1, transferL_2} 
could lead to significant speed-ups and would render the models more transferable.

\subsection{Transition state search}
Regarding the performance curve for the transition states it is encouraging to note that the
slope is substantially steeper than for the equilibrium geometry, also indicating that the OQML based energies and gradients also work well for locating saddle-points, which is unprecedented in literature, to the best of our knowledge. 
A direct one-to-one comparison to the equilibrium geometry relaxations, however, is not possible as the differences might also be due to the use of two very different optimizers (LBFGS vs.~QST2). 

\begin{figure*}[!tbp]
      \includegraphics[width=0.99\textwidth]{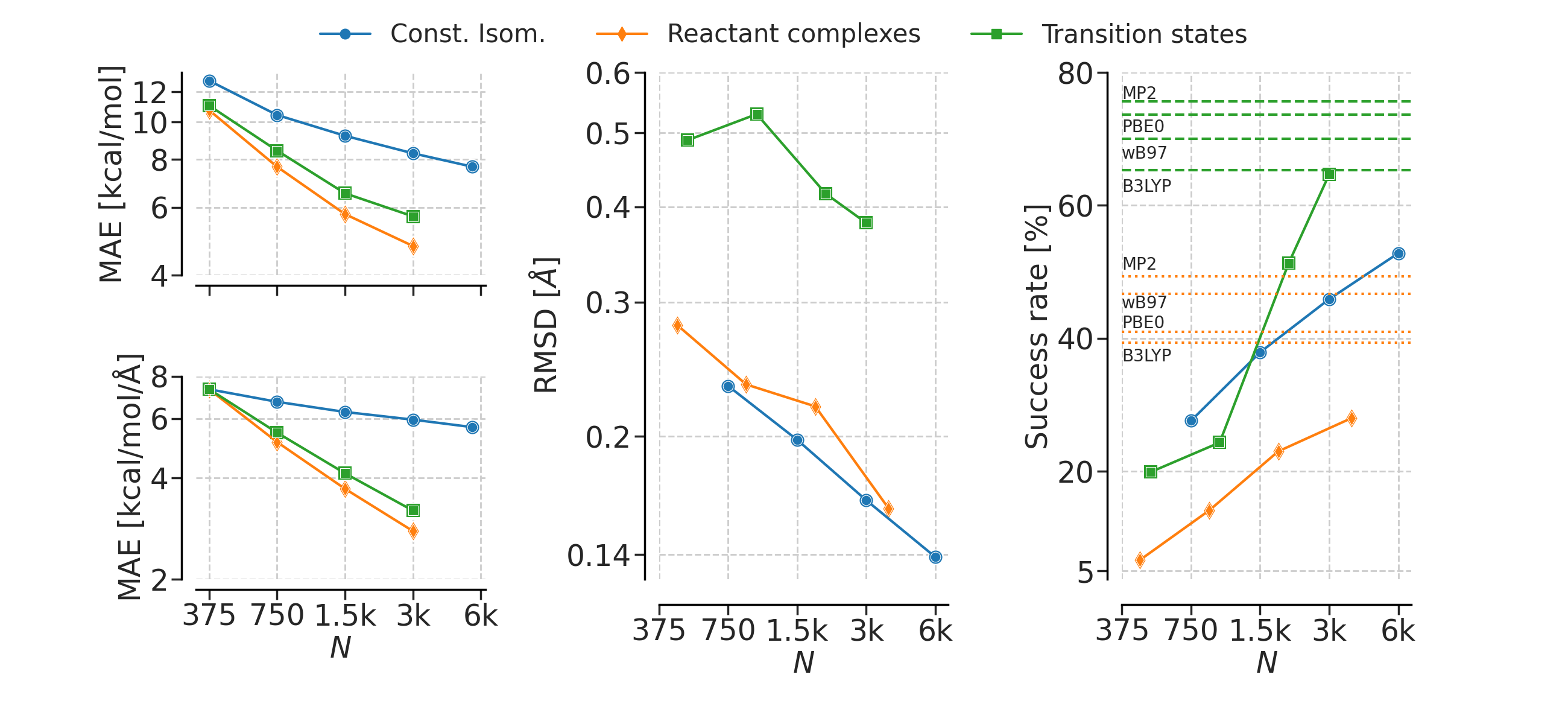}
  \caption{Learning curves for energies (top left) and atomic forces (bottom left). RMSD performance curves of geometry vs. training set size $N$ (middle). Success rates of the optimization and transition state searches (right). Colours correspond to results based on three distinct data-sets: constitutional isomers (QM9), reactant complexes (QMrxn20), and transition states (QMrxn20). Dashed green and dotted orange horizontal lines correspond to the success rate of the reference calculations for the TS searches and the geometry optimization, respectively}
  \label{fig:lc}
\end{figure*}

Performance curves for success rates have also been included in Figure \ref{fig:lc} right.
We note that for all models and data-sets the success rate of the optimization runs systematically
increases with training set size. 
We find that even for OQML models trained on small training set sizes resulting in relatively high RMSDs ($\sim$ 0.4 \r{A}), the success rate increases  from 7\% to 28\% and from 20\% to 65\% for the reactant complexes and the transition states, respectively, as shown in Figure \ref{fig:lc}, closing to the MP2 success rates (horizontal lines). 
Surprisingly, even though the RMSD performance curve for the constitutional isomer set is the best, the success performance curve is the worst. 
This could be due to the higher dimensionality in the QM9 based data-sets, where the optimizer has to locate the minimum for substantially larger systems encoding more degrees of freedom. The systematic increase in success rate, however, represents strong evidence in favor of the proposed model, as one can always improve it through mere addition of training instances, apparently resulting in increasingly smooth potential energy surfaces with fewer and fewer artifacts---an important prerequisite for successful optimization runs using algorithms such as LBFGS\cite{Liu_1989}.
%


For further analysis of reactant complexes and transition states we used 300 out-of-sample compounds.
Table \ref{tab:sum} shows a summary of the predictions of the reactant complex optimization of the ground states (GS) and the transition state (TS) search, as well as the comparison to the three DFT methods (B3LYP, PBE0, $\omega$B97X) with the 6-311G(d) basis set.
RMSDs for the geometries are around 0.1\r{A} for the DFT methods and 0.4\r{A} for the OQML method considering the transition states.
The performance of the ML model reaches the same accuracy for the reactant complexes as the DFT geometry optimization resulting in RMSD's on the order of 0.05 to 0.14~\r{A}.
Using the same model we calculated numerical frequencies and reached a mean absolute error over the 300 test transition states of 33.63 cm$^{-1}$ and 14.09 cm$^{-1}$ for transition states and reactant complexes, respectively, which is comparable to the DFT errors.

For the activation energy $E_\mathrm{a}$, the ML model reaches slightly higher MAE of 5.851 kcal/mol compared to the MAE of $\sim$1kcal/mol of the DFT methods.
Although, the error of  $\sim$6 kcal/mol is still high, other ML models could be used to learn the activation energies e.g. the R2B model\cite{Heinen2021} which was applied on this data-set and solely uses the molecular graph as input for the ML model.

We note that a direct comparison of the OQML and DFT results in Table I would not be fair as OQML was fitted on data similar to the query compounds while DFT methods and basis sets are universal in nature and were fitted against much more diverse chemistries. 

\begin{figure*}[!tbp]
      \includegraphics[width=0.99\textwidth]{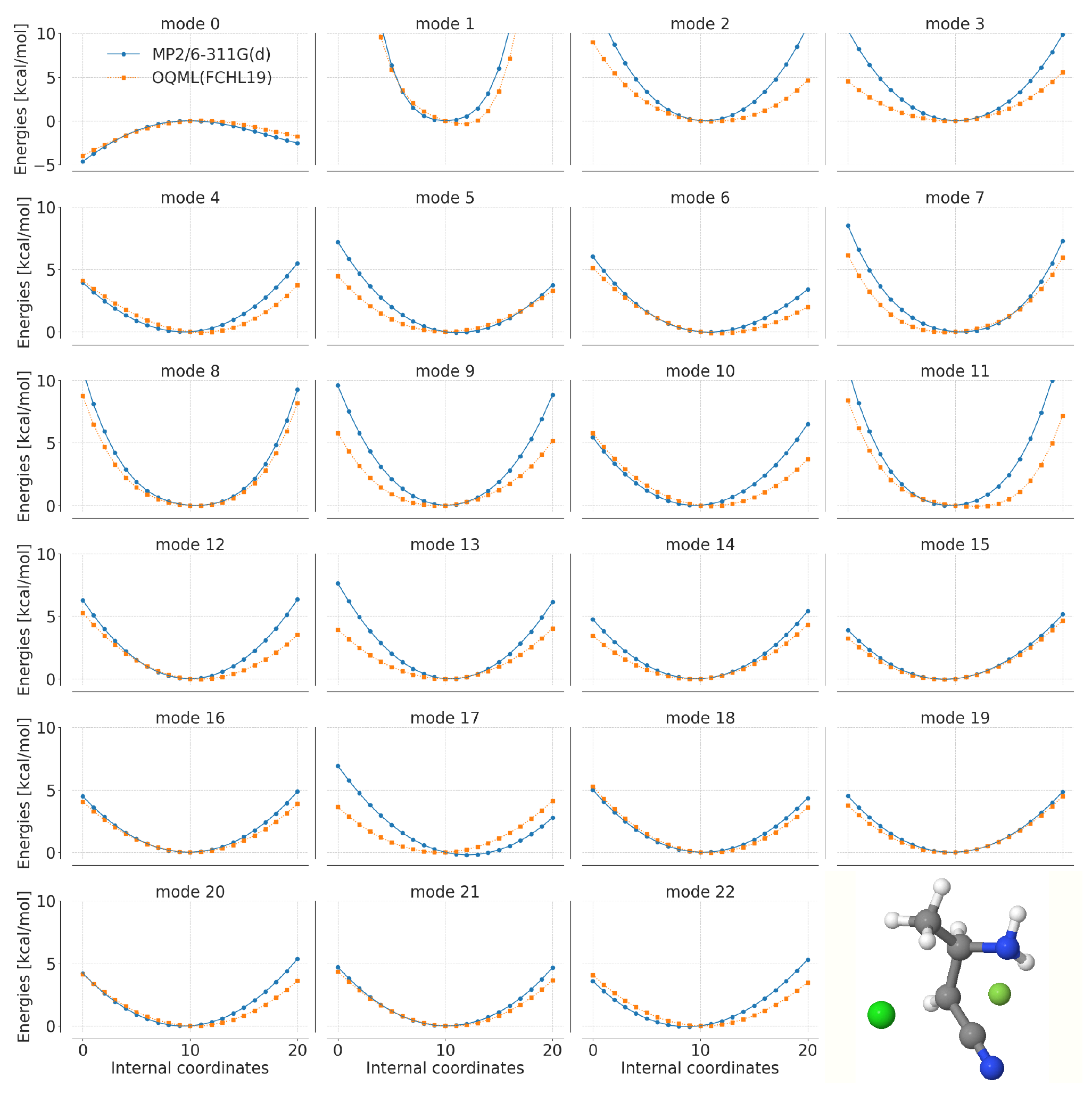}
  \caption{Example normal mode scan showing energy changes as a function of distortion along TS modes for the transition state of the S$_N$2 reaction of [H(CN)C-C(CH$_3$)(NH2)] with Cl and F as leaving group and nucleophile, respectively. Geometry of an MP2/6-311G(d) converged and validated TS was used and distorted along its normal modes. Subsequently, single point calculation (MP2), as well as ML predictions were plotted for the first 23 normal modes and their displacements. The x-axis describes the index of the distorted geometry and the y-axis describes the energy relative to the MP2 equilibrium geometry. Both energies, MP2 (blue) and ML (orange) are scaled by the equilibrium geometry (index 10).}
  \label{fig:normal_modes}
\end{figure*}

Finally, we showcase the OQML predicted results for the transition state of one randomly drawn exemplary reaction,
involving [H(CN)C-C(CH$_3$)(NH2)] with Cl and F as leaving group and nucleophile, respectively.
In Figure \ref{fig:normal_modes} the calculated transition state normal modes are shown, energies once predicted by OQML and once as obtained from MP2 for comparison.
Even though, the RMSD of the predicted geometries are off by 0.4 \r{A}, the curvature is described reasonably well by the OQML model, which is supported by the relatively small errors in frequencies, as well as by the high success rate of the transition state search.
\setlength{\tabcolsep}{14pt}
\renewcommand{\arraystretch}{1}
\begin{table*}[ht!]
\centering
\begin{tabular}{c|cc|cc|c|c} \toprule
Method & \multicolumn{2}{c}{RMSD [$\si{\angstrom}$]} & \multicolumn{2}{c}{$\Delta\nu$ [cm$^{-1}$]} & $E_{\mathrm{a}}$ [kcal$\cdot$mol$^{-1}$] & $N$\\
\toprule
 OQML (FCHL19 )              & 0.161 & 0.381 &  14.09 & 26.06 (94.21) & 5.851 & 3753/3812\\
 B3LYP/6-311G(d)             & 0.053 & 0.134 &  31.37 & 32.37 (145.18)& 1.352 & 116\cite{Lu_2015}\\
 PBE0/6-311G(d)              & 0.046 & 0.096 &  22.93 & 33.89 (147.56)& 1.016 & 106\cite{Adamo_1999,PBE01}\\
 $\omega$B97XD/6-311G(d)     & 0.033 & 0.096 &  13.94 & 33.63 (150.01)& 0.853 & 1108\cite{Mardirossian2014}\\
\toprule
\end{tabular}
\caption{Summary of results for 300 out of sample test cases for relaxation of reactants  (left) and transition state searches (right) for OQML models and three DFT methods for comparison with MP2/6-311G(d) as reference. The table shows the difference in geometry (RMSD), in frequency ($\nu$), and in activation energy ($E_\mathrm{a}$) for each method. N corresponds to the training set size of the ML models (MP2/6-311G(d)) and to the data set size for parametrization of the DFT functionals. Geometry optimizations using the LBFGS algorithm from the ASE package were truncated after 50 iterations and the default threshold (\emph{fmax}~=~0.05~eV/\r{A}) was used. The limit for the transition state search was the default iteration limit of 100 steps. Frequency values in parenthesis for the transition states are the errors of the first (imaginary frequency)}
\label{tab:sum}
\end{table*}
\section*{Conclusion}
Our findings indicate that OQML can be trained across chemical compound space in order to optimize geometries and search for saddle points (transition states) for new out-of-sample query compounds. 
As such, we have demonstrated the applicability of OQML to serve as a surrogate model of conventional quantum based energies and forces, and can be  employed within legacy optmizers.
Most notably, prediction errors of OQML in terms of RMSDs, frequencies, and activation energies systematically improve as training set sizes increase. Similarly, success rates of convergence of optimizers also improve as training set sizes grow. 

The reported learning curves exhibit linear decay as a function of  training set size on log-log scales, indicating the completeness of the model and that even further improvements in predictive power can be achieved through mere addition of further training data.
Corresponding performance curves of RMSDs suggest that the optimization process (RMSD as well as success rate) based on these models could also be further improved by increasing the training set size.
Especially for the constitutional isomers (small, organic, and closed shell molecules) the description of the potential energy surfaces improves steadily by adding more training data. 
For example, the success rate improves from 27\% to 52\% for the smallest and largest training set size, respectively.
Small deviations in predicted vibrational frequencies from reference MP2 frequencies further corroborate this point. 


In the future, the exploration of out-of-equilibrium geometries farther away from a local minima the QM7x data-set\cite{qm7x} could be investigated. 
To make OQML more transferable and applicable also to larger reactants, an Amon based extension\cite{Amons} could also be implemented.
OQML could also be helpful for the generation of larger and consistent data-sets in quantum chemistry, especially for the study of reactions.


\section*{Acknowledgement}
We thank A.~S.~Christensen for discussions. 
This project has received funding from the European Research Council (ERC) under the European Union’s Horizon 2020 research and innovation program (grant agreement No. 772834).
This result only reflects the author's view and the EU is not responsible for any use that may be made of the information it contains.
This research was also supported by the NCCR MARVEL, a National Centre of Competence in Research, funded by the Swiss National Science Foundation (grant number 182892).

%
\bibliography{literature}

\begin{thebibliography}{49}%
\makeatletter
\providecommand \@ifxundefined [1]{%
 \@ifx{#1\undefined}
}%
\providecommand \@ifnum [1]{%
 \ifnum #1\expandafter \@firstoftwo
 \else \expandafter \@secondoftwo
 \fi
}%
\providecommand \@ifx [1]{%
 \ifx #1\expandafter \@firstoftwo
 \else \expandafter \@secondoftwo
 \fi
}%
\providecommand \natexlab [1]{#1}%
\providecommand \enquote  [1]{``#1''}%
\providecommand \bibnamefont  [1]{#1}%
\providecommand \bibfnamefont [1]{#1}%
\providecommand \citenamefont [1]{#1}%
\providecommand \href@noop [0]{\@secondoftwo}%
\providecommand \href [0]{\begingroup \@sanitize@url \@href}%
\providecommand \@href[1]{\@@startlink{#1}\@@href}%
\providecommand \@@href[1]{\endgroup#1\@@endlink}%
\providecommand \@sanitize@url [0]{\catcode `\\12\catcode `\$12\catcode
  `\&12\catcode `\#12\catcode `\^12\catcode `\_12\catcode `\%12\relax}%
\providecommand \@@startlink[1]{}%
\providecommand \@@endlink[0]{}%
\providecommand \url  [0]{\begingroup\@sanitize@url \@url }%
\providecommand \@url [1]{\endgroup\@href {#1}{\urlprefix }}%
\providecommand \urlprefix  [0]{URL }%
\providecommand \Eprint [0]{\href }%
\providecommand \doibase [0]{http://dx.doi.org/}%
\providecommand \selectlanguage [0]{\@gobble}%
\providecommand \bibinfo  [0]{\@secondoftwo}%
\providecommand \bibfield  [0]{\@secondoftwo}%
\providecommand \translation [1]{[#1]}%
\providecommand \BibitemOpen [0]{}%
\providecommand \bibitemStop [0]{}%
\providecommand \bibitemNoStop [0]{.\EOS\space}%
\providecommand \EOS [0]{\spacefactor3000\relax}%
\providecommand \BibitemShut  [1]{\csname bibitem#1\endcsname}%
\let\auto@bib@innerbib\@empty
\bibitem [{\citenamefont {Strieth-Kalthoff}\ \emph {et~al.}(2020)\citenamefont
  {Strieth-Kalthoff}, \citenamefont {Sandfort}, \citenamefont {Segler},\ and\
  \citenamefont {Glorius}}]{MLreact_rev_2020}%
  \BibitemOpen
  \bibfield  {author} {\bibinfo {author} {\bibfnamefont {F.}~\bibnamefont
  {Strieth-Kalthoff}}, \bibinfo {author} {\bibfnamefont {F.}~\bibnamefont
  {Sandfort}}, \bibinfo {author} {\bibfnamefont {M.~H.~S.}\ \bibnamefont
  {Segler}}, \ and\ \bibinfo {author} {\bibfnamefont {F.}~\bibnamefont
  {Glorius}},\ }\href {\doibase 10.1039/c9cs00786e} {\bibfield  {journal}
  {\bibinfo  {journal} {Chemical Society Reviews}\ }\textbf {\bibinfo {volume}
  {49}},\ \bibinfo {pages} {6154} (\bibinfo {year} {2020})}\BibitemShut
  {NoStop}%
\bibitem [{\citenamefont {Kayala}\ \emph {et~al.}(2011)\citenamefont {Kayala},
  \citenamefont {Azencott}, \citenamefont {Chen},\ and\ \citenamefont
  {Baldi}}]{Kayala2011}%
  \BibitemOpen
  \bibfield  {author} {\bibinfo {author} {\bibfnamefont {M.~A.}\ \bibnamefont
  {Kayala}}, \bibinfo {author} {\bibfnamefont {C.-A.}\ \bibnamefont
  {Azencott}}, \bibinfo {author} {\bibfnamefont {J.~H.}\ \bibnamefont {Chen}},
  \ and\ \bibinfo {author} {\bibfnamefont {P.}~\bibnamefont {Baldi}},\ }\href
  {\doibase 10.1021/ci200207y} {\bibfield  {journal} {\bibinfo  {journal}
  {Journal of Chemical Information and Modeling}\ }\textbf {\bibinfo {volume}
  {51}},\ \bibinfo {pages} {2209} (\bibinfo {year} {2011})}\BibitemShut
  {NoStop}%
\bibitem [{\citenamefont {Wei}\ \emph {et~al.}(2016)\citenamefont {Wei},
  \citenamefont {Duvenaud},\ and\ \citenamefont {Aspuru-Guzik}}]{Wei2016}%
  \BibitemOpen
  \bibfield  {author} {\bibinfo {author} {\bibfnamefont {J.~N.}\ \bibnamefont
  {Wei}}, \bibinfo {author} {\bibfnamefont {D.}~\bibnamefont {Duvenaud}}, \
  and\ \bibinfo {author} {\bibfnamefont {A.}~\bibnamefont {Aspuru-Guzik}},\
  }\href {\doibase 10.1021/acscentsci.6b00219} {\bibfield  {journal} {\bibinfo
  {journal} {{ACS} Central Science}\ }\textbf {\bibinfo {volume} {2}},\
  \bibinfo {pages} {725} (\bibinfo {year} {2016})}\BibitemShut {NoStop}%
\bibitem [{\citenamefont {Jin}\ \emph {et~al.}(2017)\citenamefont {Jin},
  \citenamefont {Coley}, \citenamefont {Barzilay},\ and\ \citenamefont
  {Jaakkola}}]{NIPS2017_6854}%
  \BibitemOpen
  \bibfield  {author} {\bibinfo {author} {\bibfnamefont {W.}~\bibnamefont
  {Jin}}, \bibinfo {author} {\bibfnamefont {C.}~\bibnamefont {Coley}}, \bibinfo
  {author} {\bibfnamefont {R.}~\bibnamefont {Barzilay}}, \ and\ \bibinfo
  {author} {\bibfnamefont {T.}~\bibnamefont {Jaakkola}},\ }in\ \href
  {http://papers.nips.cc/paper/6854-predicting-organic-reaction-outcomes-with-weisfeiler-lehman-network.pdf}
  {\emph {\bibinfo {booktitle} {Advances in Neural Information Processing
  Systems 30}}},\ \bibinfo {editor} {edited by\ \bibinfo {editor}
  {\bibfnamefont {I.}~\bibnamefont {Guyon}}, \bibinfo {editor} {\bibfnamefont
  {U.~V.}\ \bibnamefont {Luxburg}}, \bibinfo {editor} {\bibfnamefont
  {S.}~\bibnamefont {Bengio}}, \bibinfo {editor} {\bibfnamefont
  {H.}~\bibnamefont {Wallach}}, \bibinfo {editor} {\bibfnamefont
  {R.}~\bibnamefont {Fergus}}, \bibinfo {editor} {\bibfnamefont
  {S.}~\bibnamefont {Vishwanathan}}, \ and\ \bibinfo {editor} {\bibfnamefont
  {R.}~\bibnamefont {Garnett}}}\ (\bibinfo  {publisher} {Curran Associates,
  Inc.},\ \bibinfo {year} {2017})\ pp.\ \bibinfo {pages}
  {2607--2616}\BibitemShut {NoStop}%
\bibitem [{\citenamefont {Fooshee}\ \emph {et~al.}(2018)\citenamefont
  {Fooshee}, \citenamefont {Mood}, \citenamefont {Gutman}, \citenamefont
  {Tavakoli}, \citenamefont {Urban}, \citenamefont {Liu}, \citenamefont
  {Huynh}, \citenamefont {Vranken},\ and\ \citenamefont {Baldi}}]{Fooshee2018}%
  \BibitemOpen
  \bibfield  {author} {\bibinfo {author} {\bibfnamefont {D.}~\bibnamefont
  {Fooshee}}, \bibinfo {author} {\bibfnamefont {A.}~\bibnamefont {Mood}},
  \bibinfo {author} {\bibfnamefont {E.}~\bibnamefont {Gutman}}, \bibinfo
  {author} {\bibfnamefont {M.}~\bibnamefont {Tavakoli}}, \bibinfo {author}
  {\bibfnamefont {G.}~\bibnamefont {Urban}}, \bibinfo {author} {\bibfnamefont
  {F.}~\bibnamefont {Liu}}, \bibinfo {author} {\bibfnamefont {N.}~\bibnamefont
  {Huynh}}, \bibinfo {author} {\bibfnamefont {D.~V.}\ \bibnamefont {Vranken}},
  \ and\ \bibinfo {author} {\bibfnamefont {P.}~\bibnamefont {Baldi}},\ }\href
  {\doibase 10.1039/c7me00107j} {\bibfield  {journal} {\bibinfo  {journal}
  {Molecular Systems Design {\&} Engineering}\ }\textbf {\bibinfo {volume}
  {3}},\ \bibinfo {pages} {442} (\bibinfo {year} {2018})}\BibitemShut {NoStop}%
\bibitem [{\citenamefont {Segler}\ \emph {et~al.}(2018)\citenamefont {Segler},
  \citenamefont {Preuss},\ and\ \citenamefont {Waller}}]{Segler2018}%
  \BibitemOpen
  \bibfield  {author} {\bibinfo {author} {\bibfnamefont {M.~H.~S.}\
  \bibnamefont {Segler}}, \bibinfo {author} {\bibfnamefont {M.}~\bibnamefont
  {Preuss}}, \ and\ \bibinfo {author} {\bibfnamefont {M.~P.}\ \bibnamefont
  {Waller}},\ }\href {\doibase 10.1038/nature25978} {\bibfield  {journal}
  {\bibinfo  {journal} {Nature}\ }\textbf {\bibinfo {volume} {555}},\ \bibinfo
  {pages} {604} (\bibinfo {year} {2018})}\BibitemShut {NoStop}%
\bibitem [{\citenamefont {Schwaller}\ \emph {et~al.}(2018)\citenamefont
  {Schwaller}, \citenamefont {Gaudin}, \citenamefont {L{\'{a}}nyi},
  \citenamefont {Bekas},\ and\ \citenamefont {Laino}}]{Schwaller2018}%
  \BibitemOpen
  \bibfield  {author} {\bibinfo {author} {\bibfnamefont {P.}~\bibnamefont
  {Schwaller}}, \bibinfo {author} {\bibfnamefont {T.}~\bibnamefont {Gaudin}},
  \bibinfo {author} {\bibfnamefont {D.}~\bibnamefont {L{\'{a}}nyi}}, \bibinfo
  {author} {\bibfnamefont {C.}~\bibnamefont {Bekas}}, \ and\ \bibinfo {author}
  {\bibfnamefont {T.}~\bibnamefont {Laino}},\ }\href {\doibase
  10.1039/c8sc02339e} {\bibfield  {journal} {\bibinfo  {journal} {Chemical
  Science}\ }\textbf {\bibinfo {volume} {9}},\ \bibinfo {pages} {6091}
  (\bibinfo {year} {2018})}\BibitemShut {NoStop}%
\bibitem [{\citenamefont {Lowe}(2012)}]{rxnfromlit}%
  \BibitemOpen
  \bibfield  {author} {\bibinfo {author} {\bibfnamefont {D.~M.}\ \bibnamefont
  {Lowe}},\ }\href {\doibase 10.17863/CAM.16293} {\  (\bibinfo {year} {2012}),\
  10.17863/CAM.16293}\BibitemShut {NoStop}%
\bibitem [{\citenamefont {Weiniger}(1988)}]{smiles1}%
  \BibitemOpen
  \bibfield  {author} {\bibinfo {author} {\bibfnamefont {D.}~\bibnamefont
  {Weiniger}},\ }\href@noop {} {\ \textbf {\bibinfo {volume} {28}},\ \bibinfo
  {pages} {31} (\bibinfo {year} {1988})}\BibitemShut {NoStop}%
\bibitem [{\citenamefont {D.~Weiniger}(1989)}]{smiles2}%
  \BibitemOpen
  \bibfield  {author} {\bibinfo {author} {\bibfnamefont {J.~L.~W.}\
  \bibnamefont {D.~Weiniger}, \bibfnamefont {A.~Weiniger}},\ }\href@noop {} {\
  \textbf {\bibinfo {volume} {29}},\ \bibinfo {pages} {97} (\bibinfo {year}
  {1989})}\BibitemShut {NoStop}%
\bibitem [{\citenamefont {Heinen}\ \emph {et~al.}(2020)\citenamefont {Heinen},
  \citenamefont {Schwilk}, \citenamefont {von Rudorff},\ and\ \citenamefont
  {von Lilienfeld}}]{Heinen_2020_cost_of_qm}%
  \BibitemOpen
  \bibfield  {author} {\bibinfo {author} {\bibfnamefont {S.}~\bibnamefont
  {Heinen}}, \bibinfo {author} {\bibfnamefont {M.}~\bibnamefont {Schwilk}},
  \bibinfo {author} {\bibfnamefont {G.~F.}\ \bibnamefont {von Rudorff}}, \ and\
  \bibinfo {author} {\bibfnamefont {O.~A.}\ \bibnamefont {von Lilienfeld}},\
  }\href {\doibase 10.1088/2632-2153/ab6ac4} {\bibfield  {journal} {\bibinfo
  {journal} {Machine Learning: Science and Technology}\ }\textbf {\bibinfo
  {volume} {1}},\ \bibinfo {pages} {025002} (\bibinfo {year}
  {2020})}\BibitemShut {NoStop}%
\bibitem [{\citenamefont {von Rudorff}\ \emph {et~al.}(2020)\citenamefont {von
  Rudorff}, \citenamefont {Heinen}, \citenamefont {Bragato},\ and\
  \citenamefont {von Lilienfeld}}]{QMrxn20}%
  \BibitemOpen
  \bibfield  {author} {\bibinfo {author} {\bibfnamefont {G.~F.}\ \bibnamefont
  {von Rudorff}}, \bibinfo {author} {\bibfnamefont {S.}~\bibnamefont {Heinen}},
  \bibinfo {author} {\bibfnamefont {M.}~\bibnamefont {Bragato}}, \ and\
  \bibinfo {author} {\bibfnamefont {A.}~\bibnamefont {von Lilienfeld}},\ }\href
  {\doibase 10.1088/2632-2153/aba822} {\bibfield  {journal} {\bibinfo
  {journal} {Machine Learning: Science and Technology}\ } (\bibinfo {year}
  {2020}),\ 10.1088/2632-2153/aba822}\BibitemShut {NoStop}%
\bibitem [{\citenamefont {Grambow}\ \emph {et~al.}(2020)\citenamefont
  {Grambow}, \citenamefont {Pattanaik},\ and\ \citenamefont
  {Green}}]{grambow2020}%
  \BibitemOpen
  \bibfield  {author} {\bibinfo {author} {\bibfnamefont {C.~A.}\ \bibnamefont
  {Grambow}}, \bibinfo {author} {\bibfnamefont {L.}~\bibnamefont {Pattanaik}},
  \ and\ \bibinfo {author} {\bibfnamefont {W.~H.}\ \bibnamefont {Green}},\
  }\href {\doibase 10.1038/s41597-020-0460-4} {\bibfield  {journal} {\bibinfo
  {journal} {Scientific Data}\ }\textbf {\bibinfo {volume} {7}},\ \bibinfo
  {pages} {1} (\bibinfo {year} {2020})}\BibitemShut {NoStop}%
\bibitem [{\citenamefont {Jackson}\ \emph {et~al.}(2021)\citenamefont
  {Jackson}, \citenamefont {Zhang},\ and\ \citenamefont
  {Pearson}}]{Jackson_2021}%
  \BibitemOpen
  \bibfield  {author} {\bibinfo {author} {\bibfnamefont {R.}~\bibnamefont
  {Jackson}}, \bibinfo {author} {\bibfnamefont {W.}~\bibnamefont {Zhang}}, \
  and\ \bibinfo {author} {\bibfnamefont {J.}~\bibnamefont {Pearson}},\ }\href
  {\doibase 10.1039/d1sc01206a} {\bibfield  {journal} {\bibinfo  {journal}
  {Chemical Science}\ }\textbf {\bibinfo {volume} {12}},\ \bibinfo {pages}
  {10022} (\bibinfo {year} {2021})}\BibitemShut {NoStop}%
\bibitem [{\citenamefont {Henkelman}\ \emph {et~al.}(2000)\citenamefont
  {Henkelman}, \citenamefont {Uberuaga},\ and\ \citenamefont
  {J{\'{o}}nsson}}]{Henkelman_2000}%
  \BibitemOpen
  \bibfield  {author} {\bibinfo {author} {\bibfnamefont {G.}~\bibnamefont
  {Henkelman}}, \bibinfo {author} {\bibfnamefont {B.~P.}\ \bibnamefont
  {Uberuaga}}, \ and\ \bibinfo {author} {\bibfnamefont {H.}~\bibnamefont
  {J{\'{o}}nsson}},\ }\href {\doibase 10.1063/1.1329672} {\bibfield  {journal}
  {\bibinfo  {journal} {The Journal of Chemical Physics}\ }\textbf {\bibinfo
  {volume} {113}},\ \bibinfo {pages} {9901} (\bibinfo {year}
  {2000})}\BibitemShut {NoStop}%
\bibitem [{\citenamefont {Henkelman}\ and\ \citenamefont
  {J{\'o}nsson}(2000)}]{henkelman2000improved}%
  \BibitemOpen
  \bibfield  {author} {\bibinfo {author} {\bibfnamefont {G.}~\bibnamefont
  {Henkelman}}\ and\ \bibinfo {author} {\bibfnamefont {H.}~\bibnamefont
  {J{\'o}nsson}},\ }\href {\doibase 10.1063/1.1323224} {\bibfield  {journal}
  {\bibinfo  {journal} {The Journal of chemical physics}\ }\textbf {\bibinfo
  {volume} {113}},\ \bibinfo {pages} {9978} (\bibinfo {year}
  {2000})}\BibitemShut {NoStop}%
\bibitem [{\citenamefont {Ramakrishnan}\ \emph {et~al.}(2015)\citenamefont
  {Ramakrishnan}, \citenamefont {Dral}, \citenamefont {Rupp},\ and\
  \citenamefont {von Lilienfeld}}]{deltaML2015}%
  \BibitemOpen
  \bibfield  {author} {\bibinfo {author} {\bibfnamefont {R.}~\bibnamefont
  {Ramakrishnan}}, \bibinfo {author} {\bibfnamefont {P.~O.}\ \bibnamefont
  {Dral}}, \bibinfo {author} {\bibfnamefont {M.}~\bibnamefont {Rupp}}, \ and\
  \bibinfo {author} {\bibfnamefont {O.~A.}\ \bibnamefont {von Lilienfeld}},\
  }\href {\doibase 10.1021/acs.jctc.5b00099} {\bibfield  {journal} {\bibinfo
  {journal} {Journal of Chemical Theory and Computation}\ }\textbf {\bibinfo
  {volume} {11}},\ \bibinfo {pages} {2087} (\bibinfo {year}
  {2015})}\BibitemShut {NoStop}%
\bibitem [{\citenamefont {Torres}\ \emph {et~al.}(2019)\citenamefont {Torres},
  \citenamefont {Jennings}, \citenamefont {Hansen}, \citenamefont {Boes},\ and\
  \citenamefont {Bligaard}}]{NEB_ml}%
  \BibitemOpen
  \bibfield  {author} {\bibinfo {author} {\bibfnamefont {J.~A.~G.}\
  \bibnamefont {Torres}}, \bibinfo {author} {\bibfnamefont {P.~C.}\
  \bibnamefont {Jennings}}, \bibinfo {author} {\bibfnamefont {M.~H.}\
  \bibnamefont {Hansen}}, \bibinfo {author} {\bibfnamefont {J.~R.}\
  \bibnamefont {Boes}}, \ and\ \bibinfo {author} {\bibfnamefont
  {T.}~\bibnamefont {Bligaard}},\ }\href {\doibase
  10.1103/physrevlett.122.156001} {\bibfield  {journal} {\bibinfo  {journal}
  {Physical Review Letters}\ }\textbf {\bibinfo {volume} {122}} (\bibinfo
  {year} {2019}),\ 10.1103/physrevlett.122.156001}\BibitemShut {NoStop}%
\bibitem [{\citenamefont {Mortensen}\ \emph {et~al.}(2020)\citenamefont
  {Mortensen}, \citenamefont {Meldgaard}, \citenamefont {Bisbo}, \citenamefont
  {Christiansen},\ and\ \citenamefont {Hammer}}]{asla}%
  \BibitemOpen
  \bibfield  {author} {\bibinfo {author} {\bibfnamefont {H.~L.}\ \bibnamefont
  {Mortensen}}, \bibinfo {author} {\bibfnamefont {S.~A.}\ \bibnamefont
  {Meldgaard}}, \bibinfo {author} {\bibfnamefont {M.~K.}\ \bibnamefont
  {Bisbo}}, \bibinfo {author} {\bibfnamefont {M.-P.~V.}\ \bibnamefont
  {Christiansen}}, \ and\ \bibinfo {author} {\bibfnamefont {B.}~\bibnamefont
  {Hammer}},\ }\href@noop {} {\enquote {\bibinfo {title} {Atomistic structure
  learning algorithm with surrogate energy model relaxation},}\ } (\bibinfo
  {year} {2020}),\ \Eprint {http://arxiv.org/abs/2007.07523} {arXiv:2007.07523}
  \BibitemShut {NoStop}%
\bibitem [{\citenamefont {Lemm}\ \emph {et~al.}(2021)\citenamefont {Lemm},
  \citenamefont {von Rudorff},\ and\ \citenamefont {von
  Lilienfeld}}]{Lemm_2021}%
  \BibitemOpen
  \bibfield  {author} {\bibinfo {author} {\bibfnamefont {D.}~\bibnamefont
  {Lemm}}, \bibinfo {author} {\bibfnamefont {G.~F.}\ \bibnamefont {von
  Rudorff}}, \ and\ \bibinfo {author} {\bibfnamefont {O.~A.}\ \bibnamefont {von
  Lilienfeld}},\ }\href {\doibase 10.1038/s41467-021-24525-7} {\bibfield
  {journal} {\bibinfo  {journal} {Nature Communications}\ }\textbf {\bibinfo
  {volume} {12}} (\bibinfo {year} {2021}),\
  10.1038/s41467-021-24525-7}\BibitemShut {NoStop}%
\bibitem [{\citenamefont {Meyer}\ and\ \citenamefont
  {Hauser}(2020)}]{GPR_opt1}%
  \BibitemOpen
  \bibfield  {author} {\bibinfo {author} {\bibfnamefont {R.}~\bibnamefont
  {Meyer}}\ and\ \bibinfo {author} {\bibfnamefont {A.~W.}\ \bibnamefont
  {Hauser}},\ }\href {\doibase 10.1063/1.5144603} {\bibfield  {journal}
  {\bibinfo  {journal} {The Journal of Chemical Physics}\ }\textbf {\bibinfo
  {volume} {152}},\ \bibinfo {pages} {084112} (\bibinfo {year}
  {2020})}\BibitemShut {NoStop}%
\bibitem [{\citenamefont {Born}\ and\ \citenamefont
  {Kästner}(2021)}]{GPR_opt2}%
  \BibitemOpen
  \bibfield  {author} {\bibinfo {author} {\bibfnamefont {D.}~\bibnamefont
  {Born}}\ and\ \bibinfo {author} {\bibfnamefont {J.}~\bibnamefont
  {Kästner}},\ }\href {\doibase 10.1021/acs.jctc.1c00517} {\bibfield
  {journal} {\bibinfo  {journal} {Journal of Chemical Theory and Computation}\
  } (\bibinfo {year} {2021}),\ 10.1021/acs.jctc.1c00517}\BibitemShut {NoStop}%
\bibitem [{\citenamefont {Mako{\'{s}}}\ \emph {et~al.}(2021)\citenamefont
  {Mako{\'{s}}}, \citenamefont {Verma}, \citenamefont {Larson}, \citenamefont
  {Freindorf},\ and\ \citenamefont {Kraka}}]{Mako__2021}%
  \BibitemOpen
  \bibfield  {author} {\bibinfo {author} {\bibfnamefont {M.~Z.}\ \bibnamefont
  {Mako{\'{s}}}}, \bibinfo {author} {\bibfnamefont {N.}~\bibnamefont {Verma}},
  \bibinfo {author} {\bibfnamefont {E.~C.}\ \bibnamefont {Larson}}, \bibinfo
  {author} {\bibfnamefont {M.}~\bibnamefont {Freindorf}}, \ and\ \bibinfo
  {author} {\bibfnamefont {E.}~\bibnamefont {Kraka}},\ }\href {\doibase
  10.1063/5.0055094} {\bibfield  {journal} {\bibinfo  {journal} {The Journal of
  Chemical Physics}\ }\textbf {\bibinfo {volume} {155}},\ \bibinfo {pages}
  {024116} (\bibinfo {year} {2021})}\BibitemShut {NoStop}%
\bibitem [{\citenamefont {Ramakrishnan}\ \emph {et~al.}(2014)\citenamefont
  {Ramakrishnan}, \citenamefont {Dral}, \citenamefont {Rupp},\ and\
  \citenamefont {von Lilienfeld}}]{qm9}%
  \BibitemOpen
  \bibfield  {author} {\bibinfo {author} {\bibfnamefont {R.}~\bibnamefont
  {Ramakrishnan}}, \bibinfo {author} {\bibfnamefont {P.~O.}\ \bibnamefont
  {Dral}}, \bibinfo {author} {\bibfnamefont {M.}~\bibnamefont {Rupp}}, \ and\
  \bibinfo {author} {\bibfnamefont {O.~A.}\ \bibnamefont {von Lilienfeld}},\
  }\href {\doibase 10.1038/sdata.2014.22} {\bibfield  {journal} {\bibinfo
  {journal} {Scientific Data}\ }\textbf {\bibinfo {volume} {1}} (\bibinfo
  {year} {2014}),\ 10.1038/sdata.2014.22}\BibitemShut {NoStop}%
\bibitem [{\citenamefont {Christensen}\ \emph {et~al.}(2019)\citenamefont
  {Christensen}, \citenamefont {Faber},\ and\ \citenamefont {von
  Lilienfeld}}]{Christensen2019}%
  \BibitemOpen
  \bibfield  {author} {\bibinfo {author} {\bibfnamefont {A.~S.}\ \bibnamefont
  {Christensen}}, \bibinfo {author} {\bibfnamefont {F.~A.}\ \bibnamefont
  {Faber}}, \ and\ \bibinfo {author} {\bibfnamefont {O.~A.}\ \bibnamefont {von
  Lilienfeld}},\ }\href {\doibase 10.1063/1.5053562} {\bibfield  {journal}
  {\bibinfo  {journal} {The Journal of Chemical Physics}\ }\textbf {\bibinfo
  {volume} {150}},\ \bibinfo {pages} {064105} (\bibinfo {year}
  {2019})}\BibitemShut {NoStop}%
\bibitem [{\citenamefont {Christensen}\ and\ \citenamefont {von
  Lilienfeld}(2020)}]{Christensen2020}%
  \BibitemOpen
  \bibfield  {author} {\bibinfo {author} {\bibfnamefont {A.~S.}\ \bibnamefont
  {Christensen}}\ and\ \bibinfo {author} {\bibfnamefont {O.~A.}\ \bibnamefont
  {von Lilienfeld}},\ }\href {\doibase 10.1088/2632-2153/abba6f} {\bibfield
  {journal} {\bibinfo  {journal} {Machine Learning: Science and Technology}\
  }\textbf {\bibinfo {volume} {1}},\ \bibinfo {pages} {045018} (\bibinfo {year}
  {2020})}\BibitemShut {NoStop}%
\bibitem [{\citenamefont {Faber}\ \emph {et~al.}(2018)\citenamefont {Faber},
  \citenamefont {Christensen}, \citenamefont {Huang},\ and\ \citenamefont {von
  Lilienfeld}}]{FCHL}%
  \BibitemOpen
  \bibfield  {author} {\bibinfo {author} {\bibfnamefont {F.~A.}\ \bibnamefont
  {Faber}}, \bibinfo {author} {\bibfnamefont {A.~S.}\ \bibnamefont
  {Christensen}}, \bibinfo {author} {\bibfnamefont {B.}~\bibnamefont {Huang}},
  \ and\ \bibinfo {author} {\bibfnamefont {O.~A.}\ \bibnamefont {von
  Lilienfeld}},\ }\href {\doibase 10.1063/1.5020710} {\bibfield  {journal}
  {\bibinfo  {journal} {The Journal of Chemical Physics}\ }\textbf {\bibinfo
  {volume} {148}},\ \bibinfo {pages} {241717} (\bibinfo {year}
  {2018})}\BibitemShut {NoStop}%
\bibitem [{\citenamefont {Christensen}\ \emph {et~al.}(2020)\citenamefont
  {Christensen}, \citenamefont {Bratholm}, \citenamefont {Faber},\ and\
  \citenamefont {von Lilienfeld}}]{FCHL19}%
  \BibitemOpen
  \bibfield  {author} {\bibinfo {author} {\bibfnamefont {A.~S.}\ \bibnamefont
  {Christensen}}, \bibinfo {author} {\bibfnamefont {L.~A.}\ \bibnamefont
  {Bratholm}}, \bibinfo {author} {\bibfnamefont {F.~A.}\ \bibnamefont {Faber}},
  \ and\ \bibinfo {author} {\bibfnamefont {O.~A.}\ \bibnamefont {von
  Lilienfeld}},\ }\href {\doibase 10.1063/1.5126701} {\bibfield  {journal}
  {\bibinfo  {journal} {The Journal of Chemical Physics}\ }\textbf {\bibinfo
  {volume} {152}},\ \bibinfo {pages} {044107} (\bibinfo {year}
  {2020})}\BibitemShut {NoStop}%
\bibitem [{\citenamefont {Rappe}\ \emph {et~al.}(1992)\citenamefont {Rappe},
  \citenamefont {Casewit}, \citenamefont {Colwell}, \citenamefont {Goddard},\
  and\ \citenamefont {Skiff}}]{Rappe1992}%
  \BibitemOpen
  \bibfield  {author} {\bibinfo {author} {\bibfnamefont {A.~K.}\ \bibnamefont
  {Rappe}}, \bibinfo {author} {\bibfnamefont {C.~J.}\ \bibnamefont {Casewit}},
  \bibinfo {author} {\bibfnamefont {K.~S.}\ \bibnamefont {Colwell}}, \bibinfo
  {author} {\bibfnamefont {W.~A.}\ \bibnamefont {Goddard}}, \ and\ \bibinfo
  {author} {\bibfnamefont {W.~M.}\ \bibnamefont {Skiff}},\ }\href {\doibase
  10.1021/ja00051a040} {\bibfield  {journal} {\bibinfo  {journal} {Journal of
  the American Chemical Society}\ }\textbf {\bibinfo {volume} {114}},\ \bibinfo
  {pages} {10024} (\bibinfo {year} {1992})}\BibitemShut {NoStop}%
\bibitem [{\citenamefont {Heinen}\ \emph {et~al.}(2021)\citenamefont {Heinen},
  \citenamefont {von Rudorff},\ and\ \citenamefont {von
  Lilienfeld}}]{Heinen2021}%
  \BibitemOpen
  \bibfield  {author} {\bibinfo {author} {\bibfnamefont {S.}~\bibnamefont
  {Heinen}}, \bibinfo {author} {\bibfnamefont {G.~F.}\ \bibnamefont {von
  Rudorff}}, \ and\ \bibinfo {author} {\bibfnamefont {O.~A.}\ \bibnamefont {von
  Lilienfeld}},\ }\href {\doibase 10.1063/5.0059742} {\bibfield  {journal}
  {\bibinfo  {journal} {The Journal of Chemical Physics}\ }\textbf {\bibinfo
  {volume} {155}},\ \bibinfo {pages} {064105} (\bibinfo {year}
  {2021})}\BibitemShut {NoStop}%
\bibitem [{\citenamefont {Peng}\ and\ \citenamefont {Schlegel}(1993)}]{qst2}%
  \BibitemOpen
  \bibfield  {author} {\bibinfo {author} {\bibfnamefont {C.}~\bibnamefont
  {Peng}}\ and\ \bibinfo {author} {\bibfnamefont {H.~B.}\ \bibnamefont
  {Schlegel}},\ }\href {\doibase 10.1002/ijch.199300051} {\bibfield  {journal}
  {\bibinfo  {journal} {Israel Journal of Chemistry}\ }\textbf {\bibinfo
  {volume} {33}},\ \bibinfo {pages} {449} (\bibinfo {year} {1993})}\BibitemShut
  {NoStop}%
\bibitem [{\citenamefont {Becke}(1993)}]{B3LYP_functional}%
  \BibitemOpen
  \bibfield  {author} {\bibinfo {author} {\bibfnamefont {A.~D.}\ \bibnamefont
  {Becke}},\ }\href@noop {} {\bibfield  {journal} {\bibinfo  {journal} {J.
  Chem. Phys}\ }\textbf {\bibinfo {volume} {98}},\ \bibinfo {pages} {5648}
  (\bibinfo {year} {1993})}\BibitemShut {NoStop}%
\bibitem [{\citenamefont {Stevens}\ \emph {et~al.}(1993)\citenamefont
  {Stevens}, \citenamefont {Devlin}, \citenamefont {Chabalowski},\ and\
  \citenamefont {Frisch}}]{B3LYP}%
  \BibitemOpen
  \bibfield  {author} {\bibinfo {author} {\bibfnamefont {P.~J.}\ \bibnamefont
  {Stevens}}, \bibinfo {author} {\bibfnamefont {F.~J.}\ \bibnamefont {Devlin}},
  \bibinfo {author} {\bibfnamefont {C.~F.}\ \bibnamefont {Chabalowski}}, \ and\
  \bibinfo {author} {\bibfnamefont {M.~J.}\ \bibnamefont {Frisch}},\
  }\href@noop {} {\bibfield  {journal} {\bibinfo  {journal} {J. Phys. Chem.}\
  }\textbf {\bibinfo {volume} {98}},\ \bibinfo {pages} {11623} (\bibinfo {year}
  {1993})}\BibitemShut {NoStop}%
\bibitem [{\citenamefont {Adamo}\ and\ \citenamefont
  {Barone}(1999)}]{Adamo_1999}%
  \BibitemOpen
  \bibfield  {author} {\bibinfo {author} {\bibfnamefont {C.}~\bibnamefont
  {Adamo}}\ and\ \bibinfo {author} {\bibfnamefont {V.}~\bibnamefont {Barone}},\
  }\href {\doibase 10.1063/1.478522} {\bibfield  {journal} {\bibinfo  {journal}
  {The Journal of Chemical Physics}\ }\textbf {\bibinfo {volume} {110}},\
  \bibinfo {pages} {6158} (\bibinfo {year} {1999})}\BibitemShut {NoStop}%
\bibitem [{\citenamefont {Mardirossian}\ and\ \citenamefont
  {Head-Gordon}(2014)}]{Mardirossian2014}%
  \BibitemOpen
  \bibfield  {author} {\bibinfo {author} {\bibfnamefont {N.}~\bibnamefont
  {Mardirossian}}\ and\ \bibinfo {author} {\bibfnamefont {M.}~\bibnamefont
  {Head-Gordon}},\ }\href {\doibase 10.1039/c3cp54374a} {\bibfield  {journal}
  {\bibinfo  {journal} {Physical Chemistry Chemical Physics}\ }\textbf
  {\bibinfo {volume} {16}},\ \bibinfo {pages} {9904} (\bibinfo {year}
  {2014})}\BibitemShut {NoStop}%
\bibitem [{\citenamefont {Binkley}\ \emph {et~al.}(1980)\citenamefont
  {Binkley}, \citenamefont {Pople},\ and\ \citenamefont
  {Hehre}}]{binkley1980self}%
  \BibitemOpen
  \bibfield  {author} {\bibinfo {author} {\bibfnamefont {J.~S.}\ \bibnamefont
  {Binkley}}, \bibinfo {author} {\bibfnamefont {J.~A.}\ \bibnamefont {Pople}},
  \ and\ \bibinfo {author} {\bibfnamefont {W.~J.}\ \bibnamefont {Hehre}},\
  }\href@noop {} {\bibfield  {journal} {\bibinfo  {journal} {J. Am. Chem.
  Soc.}\ }\textbf {\bibinfo {volume} {102}},\ \bibinfo {pages} {939} (\bibinfo
  {year} {1980})}\BibitemShut {NoStop}%
\bibitem [{\citenamefont {Petersson}\ \emph {et~al.}(1988)\citenamefont
  {Petersson}, \citenamefont {Bennett}, \citenamefont {Tensfeldt},
  \citenamefont {Al-Laham}, \citenamefont {Shirley},\ and\ \citenamefont
  {Mantzaris}}]{Petersson1988}%
  \BibitemOpen
  \bibfield  {author} {\bibinfo {author} {\bibfnamefont {G.~A.}\ \bibnamefont
  {Petersson}}, \bibinfo {author} {\bibfnamefont {A.}~\bibnamefont {Bennett}},
  \bibinfo {author} {\bibfnamefont {T.~G.}\ \bibnamefont {Tensfeldt}}, \bibinfo
  {author} {\bibfnamefont {M.~A.}\ \bibnamefont {Al-Laham}}, \bibinfo {author}
  {\bibfnamefont {W.~A.}\ \bibnamefont {Shirley}}, \ and\ \bibinfo {author}
  {\bibfnamefont {J.}~\bibnamefont {Mantzaris}},\ }\href {\doibase
  10.1063/1.455064} {\bibfield  {journal} {\bibinfo  {journal} {The Journal of
  Chemical Physics}\ }\textbf {\bibinfo {volume} {89}},\ \bibinfo {pages}
  {2193} (\bibinfo {year} {1988})}\BibitemShut {NoStop}%
\bibitem [{\citenamefont {Petersson}\ and\ \citenamefont
  {Al-Laham}(1991)}]{Petersson1991}%
  \BibitemOpen
  \bibfield  {author} {\bibinfo {author} {\bibfnamefont {G.~A.}\ \bibnamefont
  {Petersson}}\ and\ \bibinfo {author} {\bibfnamefont {M.~A.}\ \bibnamefont
  {Al-Laham}},\ }\href {\doibase 10.1063/1.460447} {\bibfield  {journal}
  {\bibinfo  {journal} {The Journal of Chemical Physics}\ }\textbf {\bibinfo
  {volume} {94}},\ \bibinfo {pages} {6081} (\bibinfo {year}
  {1991})}\BibitemShut {NoStop}%
\bibitem [{rms()}]{rmsd_py}%
  \BibitemOpen
  \href@noop {} {}\bibinfo {note} {Python library to calculate the RMSD,
  \texttt{https://github.com/charnley/rmsd}}\BibitemShut {NoStop}%
\bibitem [{\citenamefont {Kabsch}(1976)}]{Kabsch_1976}%
  \BibitemOpen
  \bibfield  {author} {\bibinfo {author} {\bibfnamefont {W.}~\bibnamefont
  {Kabsch}},\ }\href {\doibase 10.1107/s0567739476001873} {\bibfield  {journal}
  {\bibinfo  {journal} {Acta Crystallographica Section A}\ }\textbf {\bibinfo
  {volume} {32}},\ \bibinfo {pages} {922} (\bibinfo {year} {1976})}\BibitemShut
  {NoStop}%
\bibitem [{\citenamefont {Heinen}\ \emph {et~al.}(2022)\citenamefont {Heinen},
  \citenamefont {von Rudorff},\ and\ \citenamefont {von Lilienfeld}}]{SI}%
  \BibitemOpen
  \bibfield  {author} {\bibinfo {author} {\bibfnamefont {S.}~\bibnamefont
  {Heinen}}, \bibinfo {author} {\bibfnamefont {G.~F.}\ \bibnamefont {von
  Rudorff}}, \ and\ \bibinfo {author} {\bibfnamefont {O.~A.}\ \bibnamefont {von
  Lilienfeld}},\ }\href {\doibase 10.5281/zenodo.6823150} {\  (\bibinfo {year}
  {2022}),\ 10.5281/zenodo.6823150}\BibitemShut {NoStop}%
\bibitem [{\citenamefont {Huang}\ and\ \citenamefont {von
  Lilienfeld}(2020)}]{Amons}%
  \BibitemOpen
  \bibfield  {author} {\bibinfo {author} {\bibfnamefont {B.}~\bibnamefont
  {Huang}}\ and\ \bibinfo {author} {\bibfnamefont {O.~A.}\ \bibnamefont {von
  Lilienfeld}},\ }\href {http://arXiv.org/abs/1707.04146} {\bibfield  {journal}
  {\bibinfo  {journal} {Nature Chemistry}\ } (\bibinfo {year}
  {2020})}\BibitemShut {NoStop}%
\bibitem [{\citenamefont {Zaspel}\ \emph {et~al.}(2018)\citenamefont {Zaspel},
  \citenamefont {Huang}, \citenamefont {Harbrecht},\ and\ \citenamefont {von
  Lilienfeld}}]{multilevel}%
  \BibitemOpen
  \bibfield  {author} {\bibinfo {author} {\bibfnamefont {P.}~\bibnamefont
  {Zaspel}}, \bibinfo {author} {\bibfnamefont {B.}~\bibnamefont {Huang}},
  \bibinfo {author} {\bibfnamefont {H.}~\bibnamefont {Harbrecht}}, \ and\
  \bibinfo {author} {\bibfnamefont {O.~A.}\ \bibnamefont {von Lilienfeld}},\
  }\href {\doibase 10.1021/acs.jctc.8b00832} {\bibfield  {journal} {\bibinfo
  {journal} {Journal of Chemical Theory and Computation}\ }\textbf {\bibinfo
  {volume} {15}},\ \bibinfo {pages} {1546} (\bibinfo {year}
  {2018})}\BibitemShut {NoStop}%
\bibitem [{\citenamefont {Smith}\ \emph {et~al.}(2018)\citenamefont {Smith},
  \citenamefont {Nebgen}, \citenamefont {Zubatyuk}, \citenamefont {Lubbers},
  \citenamefont {Devereux}, \citenamefont {Barros}, \citenamefont {Tretiak},
  \citenamefont {Isayev},\ and\ \citenamefont {Roitberg}}]{transferL_1}%
  \BibitemOpen
  \bibfield  {author} {\bibinfo {author} {\bibfnamefont {J.~S.}\ \bibnamefont
  {Smith}}, \bibinfo {author} {\bibfnamefont {B.~T.}\ \bibnamefont {Nebgen}},
  \bibinfo {author} {\bibfnamefont {R.}~\bibnamefont {Zubatyuk}}, \bibinfo
  {author} {\bibfnamefont {N.}~\bibnamefont {Lubbers}}, \bibinfo {author}
  {\bibfnamefont {C.}~\bibnamefont {Devereux}}, \bibinfo {author}
  {\bibfnamefont {K.}~\bibnamefont {Barros}}, \bibinfo {author} {\bibfnamefont
  {S.}~\bibnamefont {Tretiak}}, \bibinfo {author} {\bibfnamefont
  {O.}~\bibnamefont {Isayev}}, \ and\ \bibinfo {author} {\bibfnamefont
  {A.}~\bibnamefont {Roitberg}},\ }\href {\doibase
  10.26434/chemrxiv.6744440.v1} {\  (\bibinfo {year} {2018}),\
  10.26434/chemrxiv.6744440.v1}\BibitemShut {NoStop}%
\bibitem [{\citenamefont {Smith}\ \emph {et~al.}(2019)\citenamefont {Smith},
  \citenamefont {Nebgen}, \citenamefont {Zubatyuk}, \citenamefont {Lubbers},
  \citenamefont {Devereux}, \citenamefont {Barros}, \citenamefont {Tretiak},
  \citenamefont {Isayev},\ and\ \citenamefont {Roitberg}}]{transferL_2}%
  \BibitemOpen
  \bibfield  {author} {\bibinfo {author} {\bibfnamefont {J.~S.}\ \bibnamefont
  {Smith}}, \bibinfo {author} {\bibfnamefont {B.~T.}\ \bibnamefont {Nebgen}},
  \bibinfo {author} {\bibfnamefont {R.}~\bibnamefont {Zubatyuk}}, \bibinfo
  {author} {\bibfnamefont {N.}~\bibnamefont {Lubbers}}, \bibinfo {author}
  {\bibfnamefont {C.}~\bibnamefont {Devereux}}, \bibinfo {author}
  {\bibfnamefont {K.}~\bibnamefont {Barros}}, \bibinfo {author} {\bibfnamefont
  {S.}~\bibnamefont {Tretiak}}, \bibinfo {author} {\bibfnamefont
  {O.}~\bibnamefont {Isayev}}, \ and\ \bibinfo {author} {\bibfnamefont {A.~E.}\
  \bibnamefont {Roitberg}},\ }\href {\doibase 10.1038/s41467-019-10827-4}
  {\bibfield  {journal} {\bibinfo  {journal} {Nature Communications}\ }\textbf
  {\bibinfo {volume} {10}} (\bibinfo {year} {2019}),\
  10.1038/s41467-019-10827-4}\BibitemShut {NoStop}%
\bibitem [{\citenamefont {Liu}\ and\ \citenamefont {Nocedal}(1989)}]{Liu_1989}%
  \BibitemOpen
  \bibfield  {author} {\bibinfo {author} {\bibfnamefont {D.~C.}\ \bibnamefont
  {Liu}}\ and\ \bibinfo {author} {\bibfnamefont {J.}~\bibnamefont {Nocedal}},\
  }\href {\doibase 10.1007/bf01589116} {\bibfield  {journal} {\bibinfo
  {journal} {Mathematical Programming}\ }\textbf {\bibinfo {volume} {45}},\
  \bibinfo {pages} {503} (\bibinfo {year} {1989})}\BibitemShut {NoStop}%
\bibitem [{\citenamefont {Lu}(2015)}]{Lu_2015}%
  \BibitemOpen
  \bibfield  {author} {\bibinfo {author} {\bibfnamefont {L.}~\bibnamefont
  {Lu}},\ }\href {\doibase 10.1002/qua.24876} {\bibfield  {journal} {\bibinfo
  {journal} {Int. J. Quantum Chem.}\ }\textbf {\bibinfo {volume} {115}},\
  \bibinfo {pages} {502} (\bibinfo {year} {2015})}\BibitemShut {NoStop}%
\bibitem [{\citenamefont {Ernzerhof}\ and\ \citenamefont
  {Scuseria}(1999)}]{PBE01}%
  \BibitemOpen
  \bibfield  {author} {\bibinfo {author} {\bibfnamefont {M.}~\bibnamefont
  {Ernzerhof}}\ and\ \bibinfo {author} {\bibfnamefont {G.~E.}\ \bibnamefont
  {Scuseria}},\ }\href@noop {} {\bibfield  {journal} {\bibinfo  {journal} {J.
  Comp. Phys.}\ }\textbf {\bibinfo {volume} {110}},\ \bibinfo {pages} {5029}
  (\bibinfo {year} {1999})}\BibitemShut {NoStop}%
\bibitem [{\citenamefont {Hoja}\ \emph {et~al.}(2021)\citenamefont {Hoja},
  \citenamefont {Sandonas}, \citenamefont {Ernst}, \citenamefont
  {Vazquez-Mayagoitia}, \citenamefont {DiStasio},\ and\ \citenamefont
  {Tkatchenko}}]{qm7x}%
  \BibitemOpen
  \bibfield  {author} {\bibinfo {author} {\bibfnamefont {J.}~\bibnamefont
  {Hoja}}, \bibinfo {author} {\bibfnamefont {L.~M.}\ \bibnamefont {Sandonas}},
  \bibinfo {author} {\bibfnamefont {B.~G.}\ \bibnamefont {Ernst}}, \bibinfo
  {author} {\bibfnamefont {A.}~\bibnamefont {Vazquez-Mayagoitia}}, \bibinfo
  {author} {\bibfnamefont {R.~A.}\ \bibnamefont {DiStasio}}, \ and\ \bibinfo
  {author} {\bibfnamefont {A.}~\bibnamefont {Tkatchenko}},\ }\href {\doibase
  10.1038/s41597-021-00812-2} {\bibfield  {journal} {\bibinfo  {journal}
  {Scientific Data}\ }\textbf {\bibinfo {volume} {8}} (\bibinfo {year}
  {2021}),\ 10.1038/s41597-021-00812-2}\BibitemShut {NoStop}%
\end{thebibliography}%
\end{document}